\newcommand{\beq}{\begin{equation}}
\newcommand{\eeq}{\end{equation}}
\newcommand{\bqa}{\begin{eqnarray}}
\newcommand{\eqa}{\end{eqnarray}}
\documentclass[12pt]{article}
\def\square{\vcenter{\vbox{\hrule height.4pt
          \hbox{\vrule width.4pt height8pt
          \kern8pt\vrule width.4pt}\hrule height.4pt}}}

\usepackage{epsfig}
\usepackage{psfig}
\begin{document}
\begin{flushleft}\hspace{11cm}
OHSTPY-HEP-T-97-013 \\ \hspace{11cm}
hep-ph/9709294 \\ \hspace{11cm}
\today  \\
\end{flushleft}
\vskip 10mm
\centerline{\Large\bf The Free Energy in Scalar Electrodynamics}
\centerline{\Large\bf at High Temperature}
\vskip 10mm
\centerline{Jens O. Andersen}
\centerline{\it Department of Physics, The Ohio State University,
Columbus, OH 43210}
\vskip 3mm
\begin{abstract}
{\footnotesize Massless scalar electrodynamics is studied at
high temperature
and zero chemical potential. I derive the free energy to order
$\lambda^2$, $\lambda e^2$ and $e^4$ by effective field theory methods.
The first step consists of the construction of an effective three-dimensional
field theory that is valid on distance scales $R\gg 1/T$ and whose parameters
can be written as power series in $\lambda$ and $e^2$. 
These coupling constants encode the contribution to physical quantities from 
the scale $T$.
The second step
is the use of the effective Lagrangian and perturbative calculations yield
the contributions to the free energy from the scale $eT\sim\sqrt{\lambda}T$.\\ \\
PACS number(s): 11.10.Wx, 12.20.DS}
\end{abstract}
\newpage
\pagestyle{plain}
\section{Introduction}
Quantum field theory at finite temperature have received considerably 
attention in recent years for a number of reasons.
Two major applications are phase transitions
in the early Universe and QCD at high temperature.

It was first observed by Kirzhnitz and Linde that symmetries which are 
spontaneously broken at zero temperature are normally restored at high 
temperature~\cite{kirz}.
The phase transition which has been studied most extensively is the
electroweak phase transition, which takes place when the Universe 
has a temperature of approximately 200 GeV, and where the Higgs field
acquires a nonzero expectation value.
One reason for investigating 
the electroweak phase transition is that it may be
responsible for the present baryon 
asymmetry. In order to explain the excess of baryons over antibaryons,
the electroweak phase transition must be
sufficiently strongly first order~\cite{hotewpt}. 
For realistic values of the Higgs mass, this is not the case in the
minimal standard model~\cite{hotewpt}, 
and so one is led to investigate extensions,
such as two Higgs doublet models~\cite{babe} 
and the minimal supersymmetric 
standard model~\cite{extension}.

The hadrons in QCD also undergoes a phase transition when the temperature
is raised to approximately 200 MeV. The hadrons turn into a plasma
of free quarks and gluons. This quark-gluon plasma may be produced
in heavy-ion collisions in the next generation of colliders. 
Hence, the study of QCD in the deconfined phase is very important.

It is now a well-known fact that bare or naive perturbation theory at high
temperature breaks down for soft external momenta $k$
($k\sim gT$, where $g$ is some generic coupling constant)
due to infrared divergences~\cite{pis}. 
Leading order results for physical quantities such as screening masses and
damping rates receive contributions from all orders of perturbation theory,
In order to do
consistent perturbative  calculations, it is necessary to apply the 
resummation
program of Braaten and Pisarski~\cite{pis} or extentions thereof. 
In this approach one uses effective 
propagators, and in the case of non-Abelian gauge theories one uses effective
vertices as well. This effective expansion resums the infinite subset of
diagrams which contributed to physical quantities al leading order, and hence
it is truly perturbative.

Resummation methods have been used by several authors to study the
symmetry restoration at finite temperature
of spontaneously broken gauge theories. 
Hebecker has studied the Abelian Higgs model to two-loop order 
in resummed perturbation theory~\cite{super}, 
while Fodor and Hebecker have analyzed
the electroweak phase transition, also in the 
two-loop approximation~\cite{fodor}.
The Abelian Higgs model has been used a testing ground 
for different resummation methods in hot gauge theories~\cite{kram}. 
However, it is also of
interest in its own right. since it provides a model for a relativistic
superconductor. The phase transition is then a transition from a 
superconducting phase (Higgs phase) to a normal conducting phase
(symmetric phase).
 
In some cases, as in the study of static quantities
such as the effective potential, Arnold and Espinosa have invented a
simplified resummation scheme~\cite{arnold}. 
For static quantities, it is sufficient
to dress the bosonic zero-frequency modes by thermal masses.
For all other bosonic modes as well the fermionic modes, the Matsubara
freqencies provide the necessary infrared cutoff.
Arnold and Espinosa~\cite{arnold} 
have used this resummation approach to study phase transitions
in various gauge theories.
It has also been used to compute the free energy in $g^2\Phi^4$-theory
to order $g^5$ by Parwani and Singh~\cite{parw4}
(the result to order $g^4$ was first obtained by 
Frenkel, Saa and Taylor~\cite{frenkel}). 
Corresponding calculations in QCD have been
carried out by Arnold and Zhai to order $g^4$~\cite{arnold1}, 
and extended to $g^5$
by Kastening and Zhai~\cite{kast}. 
Finally, Blaizot {\it et al}. have used the method
to compute the electric screening mass in scalar and spinor electrodynamics
to order $e^4$ and $e^5$, respectively~\cite{parw1}.

In recent years effective field theory methods have invaded the area of
high temperature field theory. 
This was initiated by the papers of Ginsparg and Landsman [15,16].
The basic idea is as follows. In the imaginary time formalism there
is a summation of Matsubara frequencies in loop diagrams. These frequencies
can be viewed as masses proportional equal to $2\pi nT$ for bosons and
$(2n+1)\pi T$ for fermions, where $n$ is positive integer. Hence, 
there is a mass hierarchy, and the fermionic modes as well as the nonzero
bosonic modes ought to decouple at distance scales $R\gg 1/T$,
analogous to the decoupling of heavy particles
at zero temperature and low energy according
to the Appelquist-Carrazone theorem~\cite{appel}.

There are several ways of constructing effective field theories at finite
temperature. At the one-loop level, one would integrate out the heavy modes.
In this context this means that the coupling constants are defined
by computing the effects of the heavy modes on static correlators.
Hence, only heavy modes circulate in loop diagrams.
Beyond one loop, one must be more careful.
It has been demonstrated by Jakov\'ac that the above procedure
generates non-local terms which cannot be expanded in local operators, and so
it is difficult to obtain 
${\cal L}_{\mbox{\footnotesize eff}}$~\cite{jakoleik}. 
The reason is 
that there can be light and heavy modes simultaneously in multi-loop diagrams,
and that one must take this into account in order to obtain the effective 
Lagrangian. 

Kajantie {\it et al.}~\cite{laine} 
have put forward an effective field theory approach,
which resolves the above problems. The coupling constants are now determined
by a matching procedure. One calculates static correlators in the full 
and in the effective theory, and demand they be the same. 
At the one-loop level, this method coincides with the original
approach to dimensional reduction~\cite{lands}.
Beyond the one-loop approximation, one omits the contributions to a correlator 
where all the modes circulating are light, but contributions
involving both heavy and ligth particles are retained. 
The effects from a loop diagram involving only static (light)
modes will be 
taken care of by the effective Lagrangian.
In many cases, one needs the correlators at zero external momentum. One
can then directly apply the effective potential, since this 
is the generator of 
one-particle irreducible Green's functions at zero external momentum. 
One advantage of using the effective potential is that this often is easier
to calculate than perorfming a direct calculation of
loop corrections to the correlators.

There is an alternative effective field theory approach due to Braaten and
Nieto~\cite{braaten}. Instead of explicitly constructing an effective theory for the
zero mode one proceeds as follows. One writes down the most general Lagrangian
in three dimensions, which respects the symmetries of the theory  
and includes all relevant degrees of freedom (i.e. no fermionic fields 
are present in ${\cal L}_{\mbox{\footnotesize eff}}$).
The coupling constants are again determined by computing static correlators
in the two theories and require that they match. 
One advantage of this approach is that we dot not explictly divide
loop corrections
to a static correlator into contributions from light and heavy modes.
This method has been applied by Braaten and Nieto 
both to $g^2\Phi^4$-theory~\cite{braaten} and QCD~\cite{braaten3}.
They have confirmed the results of 
calculations of the free energy to order $g^5$
in these theories, first obtained by resummation methods by 
Parwani and Singh~\cite{parw4}, and
by Zhai and Kastening~\cite{kast}, 
respectively. They also computed the screening mass squared
in $g^2\Phi^4$-theory to order $g^4$, which is a new result.
The present author has applied the methods to 
scalar~\cite{jens2} and spinor QED~\cite{jens}, and
reproduced the results by Blaizot {\it et al}.~\cite {parw1}
for the electric screening mass in the two theories
to order $e^4$ and $e^5$, respectively.
Similarly, the
free energy result in QED to order $e^5$ found by Zhai and Kastening
agrees with that obtained by effective field theory methods~\cite{jens}. 
This effective field theory aproach has also been used to 
calculate the screeening mass squared 
in $g^2\Phi^4$ theory to order $g^5$~\cite{jens3}.

In contrast with other theories, the literature on SQED at very high 
temperature ($T\gg T_c$) is somewhat sparse. There are no papers 
specifically devoted to the
calculation of the free energy at high temperature, although the result
to order $e^3$ and $\lambda^{3/2}$ may be easily extracted from existing
literature~\cite{super}.
In this work, I apply the effective field theory
methods of Braaten and Nieto to compute the free energy in SQED to
order $\lambda^2$ $\lambda e^2$, and $e^4$.
The result then represents the present calculational frontier.

The plan of the present paper is as follows.
In section II, we discuss dimensional reduction
and how effective field theory methods
can be applied  high temperature thermodynamics. 
In section III, the coefficients in the effective Lagrangian
are determined. In section IV we apply the effective three-dimensional
field theory to compute the free energy in scalar electrodynamics to
order $\lambda^2$, $\lambda e^2$ and $e^4$.
In section
V we summarize and conclude. In two appendices, the notation and 
conventions are given. We also list the sum-integrals
used in the full theory as well as the three-dimensional integrals 
needed in the effective theory. 

\section{Effective Field Theory and Dimensional Reduction}
In the imaginary time formalism (ITF) of quantum field theory, there
exists a path integral representation of the partition function~\cite{gins}:
\begin{equation}
{\cal Z}=\int {\cal D}\phi\exp\Big[-\int_{0}^{\beta}d\tau\int d^{3}x\,
{\cal L}_E\Big ].
\end{equation}
Here, $\phi$ is a generic field, and ${\cal L}_E$ is the Euclidean Lagrangian
which is obtained from the usual Lagrangian after Wick rotation,
$t\rightarrow -i\tau$.
Moreover, the boundary conditions in imaginary time are that
bosonic fields are
periodic in the time direction with 
period $\beta$ and that fermionic fields are
antiperiodic with the same period. If $\Phi$ denotes a bosonic field and
$\Psi$ denotes a fermionic field we have
\beq
\Phi ({\bf x},0)=\Phi ({\bf x},\tau),\hspace{1cm}
\Psi ({\bf x},0)=-\Psi ({\bf x},\tau).
\eeq
The (anti)periodicity implies that we can
decompose the fields into Fourier components characterized by their 
Matsubara frequencies:
\bqa
\label{exp}
\Phi ({\bf x},\tau)&=&\beta^{-\frac{1}{2}}\sum_{n}\phi_{n}({\bf x})
e^{i\omega_n\tau},\\
\Psi ({\bf x},\tau)&=&\beta^{-\frac{1}{2}}\sum_{n}\psi_{n}({\bf x})
e^{i\omega_n\tau}\,\,.
\eqa
Here
\bqa
\hspace{-1cm}\omega_n&=&2n\pi T,\hspace{2.05cm}\mbox{bosons}\\
\omega_n&=&(2n+1)\pi T,\hspace{1cm}\mbox{fermions},
\eqa
where $n$ is an integer. 
In four dimensions, bosonic fields are assigned
a naive scaling dimension one, while fermionics field have a naive scaling
dimension of $3/2$. This implies that the bosonic
fields in the three-dimensional
effective theory have a naive scaling dimension of $1/2$.

In the ITF, we can 
associate a free propagator $\Delta_n({\bf k})$
with the $n$th Fourier mode:
\beq
\Delta_n({\bf k})=\frac{1}{k^2+\omega_n^2}.
\eeq
Here, $k=|{\bf k}|$.
Hence, a quantum field theory at finite temperature
may be viewed as an infinite tower of fields in three dimensions, where
the Matsubara frequencies act as tree-level masses.
The $n=0$ bosonic mode is called a {\it light} or {\it static} mode, 
while the $n\neq 0$ modes
as well as the fermionic modes are termed {\it heavy} or {\it nonstatic}. 


At tree level the light modes are actually massless. If $\Phi$ denotes the
scalar field in $g^2\Phi^4$-theory,
the scalar field acquires a thermal mass of order $gT$. The thermal mass
reflects that static scalar fields are screened in the plasma, in complete
analogy to the screening of static electric fields in QED.
For the heavy modes this represents
a perturbative correction which is down by a power of the coupling, and these
modes are still characterized by a mass of order $T$. However, the light
mode is no longer massless, but its mass is of order $gT$.
Hence, we conclude that we have two widely separated mass scales at high
temperature, which are $T$ and $gT$. 
The Appelquist-Carrazone decoupling
theorem~\cite{appel} then suggests that the heavy modes decouple
on the scale $gT$. Moreover, modern developments in renormalization theory
guarantee that static correlators of the full theory
can be reproduced to any desired accuracy by an effective 
three-dimensional field theory of static mode at long distances $R\gg 1/T$
by tuning the parameters of the effective Lagrangian 
as functions of temperature and the parameters in the underlying theory.
The effective Lagrangian is
nonrenormalizable and contains infiniely many terms. 
The coupling constants of
${\cal L}_{\mbox{\footnotesize eff}}$ encode the physics on the scale $T$.
This process of going from a full four dimensional theory
to an effective three-dimensional Lagrangian is called 
{\it dimensional reduction}.
This is the key observation and the starting point for
the construction of effective field theories at finite temperature.

Let us now move on to to massless scalar electrodynamics.
The Euclidean Lagrangian of SQED is
\beq
\label{defsqed}
{\cal L}_{\mbox{\scriptsize SQED}}
=\frac{1}{4}F_{\mu\nu}F_{\mu\nu}+({\cal D_{\mu}}\Phi )
^{\dagger}({\cal D_{\mu}}\Phi )+\frac{\lambda}{6}(\Phi^{\dagger}\Phi )^2
+{\cal L}_{\mbox{\footnotesize gf}}
+{\cal L}_{\mbox{\footnotesize gh}}.
\eeq
The covariant derivative is ${\cal D}_{\mu}=\partial_{\mu}+ieA_{\mu}$. 
In the present work all calculations are carried out in the Feynman gauge, but
we emphasize that the parameters in the effective theory are gauge independent.
The gauge fixing term is 
\beq
{\cal L}_{\mbox{\footnotesize gf}}=
\frac{1}{2}(\partial_{\mu}A_{\mu})^{2},
\eeq
If $\eta$ denotes the ghost field, 
the ghost term is then 
\beq
{\cal L}_{\mbox{\footnotesize gh}}=
(\partial_{\mu}\overline{\eta})(\partial_{\mu}\eta),
\eeq
and is thus decoupled from the rest of the Lagrangian.

In $g^2\Phi^4$ theory we have seen that we have the two momentum
scales $T$ and $gT$. In scalar electrodynamics we have the scale
$T$ as well as the scale of electric screening and the scale of scalar 
screening. In the rest of the paper we shall assume that $e^2\sim\lambda\ll 1$
so that counting of powers of the coupling constants is simplified.

We call the 
effective three-dimensional field theory electrostatic scalar 
electrodynamics (ESQED), in analogy with the definitions introduced
by Braaten and Nieto in the case of QCD~\cite{braaten3}. 
The first task is to identify
the appropriate fields and the symmetries in ESQED. 
The fields can be identified (up to normalizations) with the 
zero-frequency modes of the original fields. 
${\cal L}_{\mbox{\scriptsize ESQED}}$ contains
a real massive scalar field, which we denote by $\rho$.  
This field is identified with the $n=0$ mode of the timelike component
of the gauge field in full SQED.
There is a complex massive scalar field, $\phi$, which is identifield with
$\Phi$. Finally, there is a massless gauge field, which is denoted by
$A_i^{3d}$ and is identified with the spatial component of the gauge
field in the underlying theory. 
Since the fields of ESQED are identified with the static modes of SQED,
we have to a first approximation the following relations:
\beq
\rho=\frac{1}{\sqrt{T}}A_0, \hspace{1cm}\phi=\frac{1}{\sqrt{T}}\Phi, 
\hspace{1cm}
A_i^{3d}=\frac{1}{\sqrt{T}}A_i.
\eeq

Now, ${\cal L}_{\mbox{\scriptsize ESQED}}$ must be a gauge invariant function
of the spatial fields $A_i$, up to the usual gauge fixing terms. 
This symmetry
follows from the corresponding symmetry in the full theory
and the Ward-Takahashi identity in the high temperature limit~\cite{lands}.
Since SQED is an Abelian gauge theory, there will be no 
magnetic mass~\cite{fradkin}.
The fact that the fields $\rho$ and $\Phi$ are massive reflects the
screening of static electric and static scalar fields, respectively.
There is rotational symmetry for $\rho$ as well
a $Z_2$-symmetry for the fields $\phi$ and $\rho$. 
The effective Lagrangian then has the general form
\bqa \nonumber
\label{defleff}
{\cal L}_{\mbox{\scriptsize ESQED}}&=&
\frac{1}{4}F_{ij}F_{ij}+({\cal D}_{i}\phi )^{\dagger}
({\cal D}_{i}\phi )+M^{2}(\Lambda )
\phi^{\dagger}\phi
+\frac{1}{2}(\partial_{i}\rho)^2+\\
&&
\frac{1}{2}m^{2}(\Lambda )\rho^{2}
+h^{2}_{E}(\Lambda)\phi^{\dagger}\phi\rho^2 
+\frac{\lambda_3}{6}(\phi^{\dagger}\phi )^2
+{\cal L}_{\mbox{\footnotesize gf}}
+{\cal L}_{\mbox{\footnotesize gh}}
+\delta{\cal L}.
\eqa
Here, $\Lambda$ is the ultraviolet cutoff in the effective theory.
The $\Lambda$-dependence of the parameters is necessary in order to cancel
the dependence of $\Lambda$ which arises in perturbative calculations in
electrostatic scalar electrodynamics.
The scale $\Lambda$ can be thought of as an arbitrary factorization scale
separating the scale $T$ and $eT$.
Furthermore, $\delta {\cal L}$ represents all local 
terms that can be constructed out of $A_{i}$ and $\rho$, which 
respect
the symmetries of the theory. This 
includes 
renormalizable terms, such as $g_{E}(\Lambda )\rho^{6}$, as well 
as 
nonrenormalizable ones like $h_{E}(\Lambda )(F_{ij}F_{ij})^{2}$.

It is not too difficult to find the order in the couplings at which
a certain operator in 
${\cal L}_{\mbox{\footnotesize eff}}$ starts to contribute to a physical 
quantity. Take e.g. $\lambda_E(\Lambda )\rho^{4}$; The coefficient in front
of the quartic coupling at leading order in $e^2$, 
can be found by considering the one-loop contribution
to the four-point function for timelike photons at zero external momenta.
Thus, it goes like $e^4$. Moreover, this operator contributes to the
free energy first at the two-loop level (the double bubble), 
and each loop is proportional
to $m_E(\Lambda)$. Since the mass goes like $eT$, this
implies that the leading order contribution is of order $e^6$.
Other operators are analyzed in a similar manner.

In~(\ref{defleff}), we did not include the unit operator. The coefficient
of the unit operator, which we denote $f_{\mbox{\scriptsize ESQED}}(\Lambda)$, 
gives the contribution
to the free energy from the momentum scale $T$. So if we are interested
in calculating the pressure we must determine it, as we determine other
coefficients in ESQED. 
Generally, this 
coefficient also 
depends on the renormalization scale $\Lambda$. By including 
$f_{\mbox{\scriptsize ESQED}}(\Lambda )$ in 
${\cal L}_{\mbox{\scriptsize ESQED}}$, we have two equivalent ways of
writing the partition function in SQED in terms of its path integral 
representation. In the full theory we have
\begin{equation}
\label{z1}
{\cal Z}=\int {\cal D}\overline{\eta}\,{\cal D}
\eta\,{\cal D}A_{\mu}\,{\cal D}\overline{\psi}
\,{\cal D}
\psi \exp\Big[-\int_{0}^{\beta}d\tau\int d^{3}x\,
{\cal L}\Big ],
\end{equation}
where $\eta$ denotes the ghost field.
The result using the effective three-dimensional theory is
\begin{equation}
\label{z2}
{\cal Z}=e^{-f_{\mbox{\scriptsize ESQED}}(\Lambda )V}
\int {\cal D}\overline{\eta}
\,{\cal D}\eta\,{\cal D}A_{i}\,{\cal D}
\rho\exp
\Big[-\int d^{3}x\,{\cal L}_{\mbox{\scriptsize ESQED}}\Big].
\end{equation}
The free energy may then be written as the sum of two terms:
\beq
\label{twoterm}
{\cal F}=T\Big[f_{\mbox{\scriptsize ESQED}}(\Lambda)-
\ln{\cal Z}_{\mbox{\scriptsize ESQED}}/V\Big],
\eeq
The first term represents the contribution
from the scale $T$ and the second term represents the contribution
from the scale $eT$, and may be calculated perturbatively using ESQED.

Let us close this section by pointing out a difference between
Abelian and Non-Abelian gauge theories. 
In QCD, we have three momentum scales. The scale $T$, which is a typical
momentum of a particle, the scale $gT$ which is the scale of colour
electric screening, and $g^2T$, which is the scale of colour magnetic
screening. So, it is convenient to construct a sequence of two effective field
theories in order to unravel the contribution to the free energy
from the three momentum scales $T$, $gT$ and $g^2T$, as
demonstrated by Braaten and Nieto~\cite{braaten3}.
Since the gauge field in ESQED is massless, one could construct a second
effective field theory only involving $A_i^{3d}$ by integrating out
the massive fields $\rho$ and $\Phi$. However, below we shall argue that
this is in fact not necessary.

We call this effective field theory magnetostatic scalar 
electrodynamics, (MSQED), and it is valid on distance scales $R\gg 1/eT$.
The Lagrangian 
consists of all terms made up by $A_i^{3d}$ which respect gauge invariance
and the other symmetries of MSQED. At leading order in the couplings
of full SQED, MSQED is simply free Maxwell theory in three dimensions. 
However, the only ultraviolet divergences one 
would encounter doing perturbative
calculations in MSQED are powerlike. The coefficients in front of these
depend upon the regulator, and are thus artifacts of the 
regulator
Because of the unphysical character of these divergences,
it is convenient to use a regulator
where the these divergences are subtracted. Dimensional regularization
provides such a procedure, since power divergences by definition
are set to zero.
This implies that there will be no contribution to the free energy
from perturbative calculations in MSQED.
This fact reflects that there is no magnetic screening in Abelian gauge 
theories and no contribution from the scale $e^2T$ to the free energy.
Hence, there is no need to construct MSQED. 
\section{The Coefficients in the Effective Lagrangian}
In this section we shall determine the parameters 
of ${\cal L}_{\mbox{\scriptsize ESQED}}$ to the order in $\lambda$
and $e^2$ 
which is required for obtaining the free  energy to order
$\lambda^2$, $\lambda e^2$ and $e^4$.
We know from the previous discussion that static correlators in SQED
at long distances $R\gg 1/T$ can be reproduced
in ESQED by tuning the parameters as functions of $\lambda$, $e^2$, $T$
and the renormalization scale $\Lambda$.
We determine the coefficients in 
${\cal L}_{\mbox{\scriptsize ESQED}}$ by the use of 
{\it strict perturbation theory}. This way of determining the
parameters of the effective three-dimensional theories were first carried
out by Braaten and Nieto in Ref.~\cite{braaten}.
Strict perturbation theory is simply ordinary perturbation
theory and is therefore plagued with infrared divergences, which become more 
and more severe as we go to higher orders in the loop expansion.
Accordingly, we split the Lagrangian of SQED into 
a free piece and an interacting piece:
\bqa \nonumber
\label{strict1}
({\cal L}_{\mbox{\scriptsize SQED}})_{0}&=&\frac{1}{4}
F_{\mu\nu}F_{\mu\nu}+(\partial_{\mu}
\Phi )^{\dagger}(\partial_{\mu}\Phi )
+{\cal L}_{\mbox{\footnotesize gf}}
+{\cal L}_{\mbox{\footnotesize gh}},
\\ 
({\cal L}_{\mbox{\scriptsize SQED}})
_{\mbox{\footnotesize int}}&=&
e^{2}\Phi^{\dagger}\Phi A_{\mu}^{2}
-ieA_{\mu}
(\Phi^{\dagger}\partial_{\mu}\Phi-\Phi\partial_{\mu}
\Phi^{\dagger})
+\frac{\lambda}{6}(\Phi^{\dagger}\Phi)^2.
\eqa
Although strict perturbation theory breaks down on the scale $eT$, we can
nevertheless use it as a tool for determining the parameters of ESQED.
The idea is that the coefficient of the effective Lagrangian  are sensitive
to the scale $T$, but insensitive to the scale $eT$.
We only have to be sure that we make the same incorrect assumptions
using the effective theory in matching calculations.
The infrared divergences which occur in 
perturbative calculations in the effective theory
will then be exactly the same as those encountered in the full theory,
and so they cancel against each other in the matching procedure.
In ESQED, these incorrect assumptions amounts to treating the mass
parameters as interactions.
The partition of the Lagrangian of ESQED is 
\bqa \nonumber
\label{strict2}
({\cal L}_{\mbox{\scriptsize ESQED}})_{0}&=&\frac{1}{4}
F_{ij}F_{ij}+(\partial_{i}
\phi )^{\dagger}(\partial_{i}\phi )
+\frac{1}{2}(\partial_{i}\rho)^{2}
+{\cal L}_{\mbox{\footnotesize gf}}
+{\cal L}_{\mbox{\footnotesize gh}},
\\ \nonumber
({\cal L}_{\mbox{\scriptsize ESQED}})
_{\mbox{\footnotesize int}}&=&\frac{1}{2}m^{2}_E(\Lambda )\rho^{2}+
M^{2}(\Lambda )\phi^{\dagger}
\phi+
e^{2}_{E}(\Lambda)\phi^{\dagger}\phi A_{i}^{3d}A_{i}^{3d}+
h^2_E(\Lambda)\phi^{\dagger}\phi\rho^{2}\\
&&-ie_{E}(\Lambda)A_{i}^{3d}
(\phi^{\dagger}\partial_{i}\phi-\phi\partial_{i}
\phi^{\dagger})+\frac{\lambda_3(\Lambda)}{6}(\phi^{\dagger}\phi)^2+\delta{\cal L}.
\eqa
Since the strict pertubative expansion given by~(\ref{strict1})
and~(\ref{strict2}) is naive perturbation theory, it 
should be clear that the
coupling constants of ESQED can be written as power series in $\lambda$ and
$e^2$. 

In full SQED, wiggly, dashes and dotted lines denote the propagators
of photons, charged scalars and ghosts, respectively.
In ESQED, the same conventions apply. Moreover,
solid lines denote the propagators
of the real scalar field $\rho$.
\subsection{Coupling Constants}
In the present work, one needs the coupling constants 
$\lambda_3^{2}(\Lambda)$, $e_{\footnotesize E}^2(\Lambda )$
and $h_{\footnotesize E}^2(\Lambda )$
to leading order in 
$\lambda$ and $e^2$. At this order one can simply
read off these coefficients from the Lagrangian in full SQED.
We have previously found that 
$\rho=\frac{1}{\sqrt{T}}A_0$ , $\phi=\frac{1}{\sqrt{T}}\Phi$ 
and $A_i^{3d}=\frac{1}{\sqrt{T}}A_i$
at leading order.
Substituting these expressions into~(\ref{defsqed})
and comparing it with~(\ref{defleff}), 
we conclude that
\beq
\lambda_3(\Lambda)=\lambda T.
\eeq
Similarly, one finds
\beq
e_{\footnotesize E}^{2}(\Lambda)=e^2 T,\hspace{1cm}
h_{\footnotesize E}^{2}(\Lambda)=e^2 T.
\eeq
There is no dependence on the renormalization scale $\Lambda$.

The above parameters have been obtained at next-to-leading order 
in $\lambda$  and $e^2$
in e.g. Ref.~\cite{peisa} by considering the appropriate static correlators. 
\subsection{Mass parameters}
We shall now determine the mass parameters $m^2_E(\Lambda )$
and $M^2(\Lambda )$ appearing in the effective Lagrangian. These parameters
are needed to leading order in $\lambda$ and $e^2$.
The mass parameter $m_E^2(\Lambda)$ has previously
been calculated by the present
author in the two-loop approximation using strict perturbation 
theory~\cite{jens2}. 
The mass parameter $M^2(\Lambda)$
has also been calculated to two-loop order
in Ref.~\cite{peisa} using the effective potential.
Nevertheless, we include the calculations here for completeness.
 
The mass parameters have the following
interpretation; they represent the contribution from the scale $T$ to 
the electric and scalar screening mass, respectively.
These  screening masses give information of the screening of  
static electric and scalar fields in the plasma~\cite{bellac}.
Now, the scalar screenning mass cannot be calculated beyond leading
order due to a mass shell singularity,
as discussed by Blaizot and Iancu in Ref.~\cite{nonp}.
The problem is the same as in QCD, namely a scalar field
interacting with a massless gauge field in three dimensions.
In QCD, this singularity may be screened by a magnetic mass 
of nonperturbative origin. In SQED the magnetic mass is absent
since it is an Abelian theory~\cite{fradkin}, and so
the problem cannot be solved this way.
We shall not discuss this any further, but refer to Ref.~\cite{nonp}
where a nonperturbative definition of the scalar screening mass is discussed
in detail.
Nevertheless, these infrared problems do not 
prevent us from determining the mass parameter
$M^2(\Lambda)$ using the strict perturbation expansion.

There are several ways of determining the mass parameters, and the simplest way
to do so is to match the screening masses in the two theories. The screening
mass is defined as the pole position of the propagator at spacelike 
momentum~\cite{braaten}. 
Let us denote the static self-energy 
function of the zeroth
component of the gauge field by $\Pi({\bf k})$. Similarly, we denote
the static scalar self-energy function by
$\Sigma({\bf k})$.
The corresponding screening masses are $m^2_s$ and $M^2_s$, respectively.
In the full theory, one then finds that the screning masses are solutions
of the following set of equations:
\bqa
\label{full}
k^2+\Pi^{}({\bf k})=0,\hspace{1cm}k^2=-m^2_{s},\\
\label{full2}
k^2+\Sigma^{}({\bf k})=0,\hspace{1cm}k^2=-M^2_{s}.
\eqa
The self-energy functions can be expanded in powers of the external momentum
${\bf k}$ and also in the number of loops in the loop expansion.
The nth order contribution to the self-enregy functions are then
denoted by $\Pi^{(n)}({\bf k})$ and $\Sigma^{(n)}({\bf k})$, respectively.
At leading order in the couplings, we simply substitute the 
one-loop approximations to the self-energy functions
into~(\ref{full}) and~(\ref{full2}). The solutions are
\beq
m^2_s\approx\Pi^{(1)}(0),\hspace{1cm}M^2_s\approx\Sigma^{(1)}(0).
\eeq
The symbol $\approx$ is a reminder that the expressions
are obtained using strict perturbation theory.
Since strict perturbation theory treat infrared
effects incorrectly, the mass parameters generally
do not coincide with the physical screening masses~\cite{braaten}. 
At leading order, however, they normally do coincide.

\begin{figure}[htb]
\begin{center}
\mbox{\psfig{figure=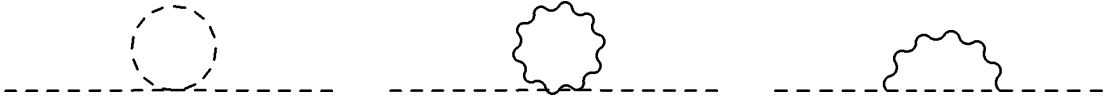}}
\end{center}
\caption[One-loop scalar self-energy diagrams in SQED.]{\protect One-loop scalar self-energy diagrams in SQED.}
\label{1skalar}
\end{figure}

\begin{figure}[htb]
\begin{center}
\mbox{\psfig{figure=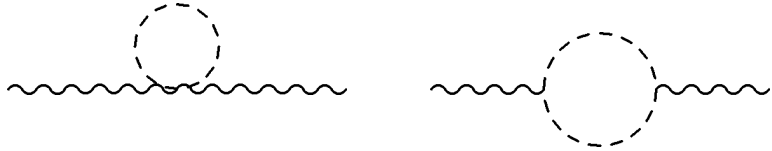}}
\end{center}
\caption[One-loop diagrams for the photon 
polarization tensor in the full theory.]{\protect One-loop self-energy diagrams in the full theory.}
\label{1lsqedm}
\end{figure}

In the full theory the Feynman graphs in Figs.~\ref{1skalar} 
and~\ref{1lsqedm} 
display the contributions at the one-loop level
to the the self-energy function of $\rho$ and $\Phi$, respectively.
We find
\bqa
\label{skmass}
\Pi^{(1)}(0)&\approx&2e^{2}\hbox{$\sum$}\!\!\!\!\!\!\int_{P}\frac{1}{P^{2}}
-4e^{2}\hbox{$\sum$}\!\!\!\!\!\!\int_{P}\frac{p^{2}_{0}}{P^{4}}, \\
\Sigma^{(1)}(0)&\approx&(d-1)e^{2}\hbox{$\sum$}\!\!\!\!\!\!\int_{P}\frac{1}{P^{2}}
+\frac{2\lambda}{3}\hbox{$\sum$}\!\!\!\!\!\!\int_{P}\frac{1}{P^{2}}.
\eqa
Here, $d=4-2\epsilon$.
In ESQED, the self-energy function of the real scalar field $\rho$ is
denoted by $\Pi_{\mbox{\footnotesize eff}}({\bf k},\Lambda)$,
while the corresponding self-energy function of $\phi$ is
$\Sigma_{\mbox{\footnotesize eff}}({\bf k},\Lambda)$.
In the effective theory the screening masses are by the following set of
equations, where the superscript ``(1)'' indicates that the
self-energies are taken in the one-loop approximation:
\bqa
\label{eff}
k^2+m^2(\Lambda )+\Pi^{(1)}_{\mbox{\footnotesize eff}}(0,\Lambda)=0,\hspace{1cm}k^2=-m^2_{s},\\
k^2+M^2(\Lambda )+\Sigma^{(1)}_{\mbox{\footnotesize eff}}(0,\Lambda)=0,
\hspace{1cm}k^2=-M^2_{s}.
\eqa
In the effective theory, all diagrams vansih identically;
Since all fields are massless in strict perturbation theory, there is no scale
in the loop integrals when the external momentum ${\bf k}$ is set to zero.  
The integrals are set to zero in dimensional regularization. Hence, 
$\Pi^{(1)}_{\mbox{\footnotesize eff}}(0,\Lambda)=0$ and 
$\Sigma^{(1)}_{\mbox{\footnotesize eff}}(0,\Lambda)=0$
The matching conditions then imply that
$m^2_E(\Lambda )=\Pi^{(1)}(0)$ and $M^{2}(\Lambda )=\Sigma^{(1)}(0)$.
Using~(\ref{1l1})$-$(\ref{1l3}) of Appendix A, we obtain
the mass parameters at leading order in $\lambda$ and $e^2$: 
\bqa
m^2_E(\Lambda )&=&\frac{e^2T^2}{3},\\
M^{2}(\Lambda )&=&\frac{e^2T^2}{4}+\frac{\lambda T^2}{18}.
\eqa
To leading order in the couplings, then, the
mass parameters are independent of $\Lambda$.

\subsection{Coefficient of the Unit Operator}
\begin{figure}[htb]
\begin{center}
\mbox{\psfig{figure=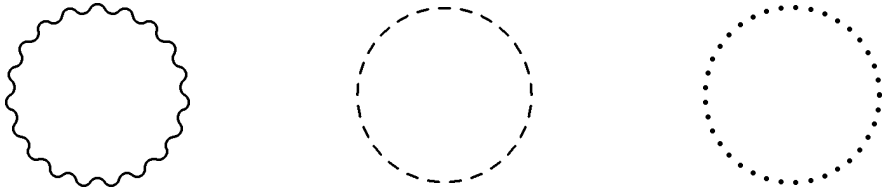}}
\end{center}
\caption{\protect One-loop vacuum diagrams in SQED.}
\label{1lsqed}
\end{figure}

In this subsection we shall determine the coefficient of the
unit operator to next-to-leading order in $\lambda$ and $e^2$. 
We shall consider the one, two and three-loop contributions separately. 
The matching condition we use to determine 
$f_{\mbox{\scriptsize ESQED}}(\Lambda)$ 
follows
from the two path integral representations of the partition function
\begin{equation}
\ln {\cal Z}=-f_{\mbox{\scriptsize ESQED}}(\Lambda )V+\ln 
{\cal Z}_{\mbox{\scriptsize ESQED}}.
\end{equation}
Let us first focus on the SQED. 
The one-loop graphs in the underlying theory are depicted in Fig.~\ref{1lsqed}
and the corresponding contribution
reads
\beq
\label{1full}
\frac{d}{2}\hbox{$\sum$}\!\!\!\!\!\!\int_{P}\ln P^2=-\frac{2\pi^2}{45}T^4.
\eeq
This is simply the free energy of a noninteracting gas of photons and
charged scalars.

\begin{figure}[htb]
\begin{center}
\mbox{\psfig{figure=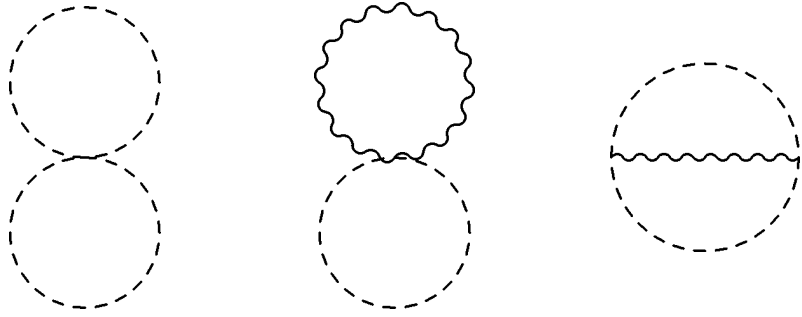}}
\end{center}
\caption{\protect Two-loop diagrams for the free energy in SQED.}
\label{2lsqed}
\end{figure}

The two-loop diagrams are shown Fig.~\ref{2lsqed}. 
After some purely algebraic manipulations,
they factorize into products of simpler one-loop sum-integrals. Using
Appendix A, we obtain
\beq
\label{toloopfull}
\frac{Z_{\lambda}\lambda}{3}\hbox{$\sum$}\!\!\!\!\!\!\int_{PQ}\frac{1}{P^2Q^2}
+(d-\frac{3}{2})Z_{e^2}e^2\hbox{$\sum$}\!\!\!\!\!\!\int_{PQ}\frac{1}{P^2Q^2}
=\Big(\frac{T^2}{12}\Big)^2\Big[\frac{\lambda}{3}+\frac{5e^2}{2}\Big].
\eeq
Here, $Z_{\lambda}$ and $Z_{e^2}$ are the renormalization constants for the
quartic coupling and the gauge coupling, respectively.
 
\begin{figure}[htb]
\begin{center}
\mbox{\psfig{figure=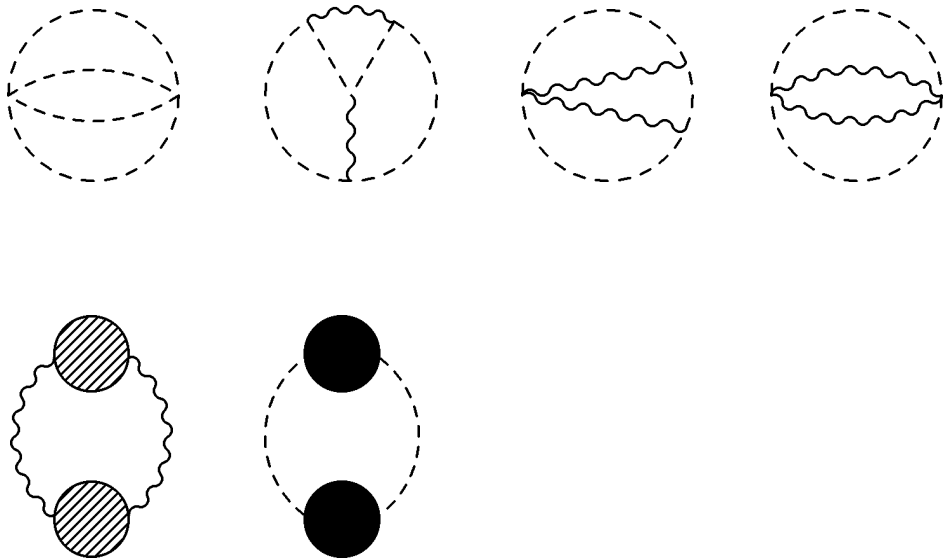}}
\end{center}
\caption{\protect Three-loop diagrams contributing to the free energy
in the underlying theory.}
\label{sqed3}
\end{figure}

Let us now turn to the three-loop diagrams. These are displayed in 
Fig.~\ref{sqed3}. Here, the shaded blob means insertion of the 
one-loop photon polarization
tensor $\Pi_{\mu\nu}(k_0,{\bf k})$, while the black blob implies insertion of the
scalar self-energy function $\Sigma (k_0,{\bf k})$, also in the one-loop
approximation.

The first four diagrams can expressed entirely in terms of the bosonic
basketball, which is defined in Appendix A. 
After some purely algebraic manipulations involving several changes of 
variables, one finds:
\beq
-\Big[\frac{\lambda^2}{18}+(d-13/4)e^4\Big]
\hbox{$\sum$}\!\!\!\!\!\!\int_{PQK}\frac{1}{P^2Q^2K^2(P+Q+K)^2}.
\eeq
The fifth diagram gives a contribution
\bqa\nonumber
-\frac{1}{4}
\hbox{$\sum$}\!\!\!\!\!\!\int_{P}\frac{1}{P^4}\Big[\Pi_{\mu\nu}(p_0,{\bf p})
\Big]^2&=&
\frac{e^4}{4}\hbox{$\sum$}\!\!\!\!\!\!\int_{PQK}\frac{1}{P^2Q^2K^2(P+Q+K)^2}\\
\nonumber
&&
-e^4\hbox{$\sum$}\!\!\!\!\!\!\int_{PQK}\frac{(K-Q)^4}{P^4Q^2K^2(P-K)^2(P-Q)^2}\\\
&&
-(d-6)e^4\hbox{$\sum$}\!\!\!\!\!\!\int_{PQK}\frac{1}{P^4Q^2K^2}.
\eqa
The last graph reads
\bqa\nonumber
\label{reveal}
-\frac{1}{2}
\hbox{$\sum$}\!\!\!\!\!\!\int_{P}\frac{1}{P^4}\Big[\Sigma (p_0,{\bf p})\Big]^2&=&
-\Big[\frac{2\lambda^2}{9}+\frac{2(d-1)\lambda e^2}{3}
+\frac{(d-1)^2e^4}{2}\Big]\hbox{$\sum$}\!\!\!\!\!\!\int_{PQK}
\frac{1}{P^4Q^2K^2}\\
&&
-2e^4\hbox{$\sum$}\!\!\!\!\!\!\int_{PQK}\frac{1}{P^2Q^2K^2(P+Q+K)^2}.
\eqa
Note that the fifth as well as the sixth diagram are infrared divergent.
It stems from the facts that both the zeroth component of the 
static photon polarization tensor, $\Pi_{00}(0,{\bf k})$,
and the static scalar self-energy function, $\Sigma (0,{\bf k})$,
are nonzero in the infrared limit ${\bf k}\rightarrow 0$.
These diagrams are the first in an infinite series of infrared graphs
which are called ring diagrams or plasmon diagrams. They are obtained
from the two diagrams considered above, by one or more insertions
of $\Pi_{\mu\nu}(k_0,{\bf k})$ and $\Sigma (k_0,{\bf k})$, respectively.

Renormalization of the quartic coupling is carried out by
substituting $Z_{\lambda}$ into~(\ref{toloopfull}). To order
$\lambda^2$, $\lambda e^2$ and $e^4$, the renormalization constant reads
\beq
Z_{\lambda}\lambda=\lambda
+\frac{5\lambda^2-18\lambda e^2+54e^4}{48\pi^2\epsilon}.
\eeq
Similarly, we renormalize the gauge coupling by subtituting
$Z_{e^2}$ into~(\ref{toloopfull}). To order $e^4$, 
$Z_{e^2}e^2$ is given by
\beq
Z_{e^2}=e^2+\frac{e^{4}}{48\pi^2\epsilon},
\eeq
The result in full SQED is now given by the sum of~\ref{1full}-~\ref{reveal}.
After we have carried out coupling constant renormalization, we are 
still left with terms proportional to $1/\epsilon$:
\bqa\nonumber
\label{unit1}
\!\!\!\!\!\!\!\!\hspace{-0.5cm}
-\frac{T\ln {\cal Z}}{V}&\approx&
-\frac{2\pi^2T^4}{45}+\Big(\frac{T^2}{12}\Big)^2\Big[\frac{\lambda}{3}
+\frac{5e^2}{2}\Big]\\ \nonumber
&&-\frac{\lambda^2}{16\pi^2}\Big(\frac{T^2}{12}\Big)^2
\Big[\frac{10}{9}\ln\frac{\Lambda}{4\pi T}+\frac{4}{9}\gamma_{E}
+\frac{31}{45}
-\frac{2}{3}\frac{\zeta^{\prime}(-3)}{\zeta (-3)}
+\frac{4}{3}\frac{\zeta^{\prime}(-1)}{\zeta (-1)}
\Big]\\ \nonumber
&&-\frac{\lambda e^2}{16\pi^2}\Big(\frac{T^2}{12}\Big)^2
\Big[\frac{4}{\epsilon}+20\ln\frac{\Lambda}{4\pi T}+4\gamma_{E}+\frac{44}{3}
+16\frac{\zeta^{\prime}(-1)}{\zeta (-1)}
\Big]\\ \nonumber
&&-\frac{e^4}{16\pi^2}\Big(\frac{T^2}{12}\Big)^2
\Big[\frac{18}{\epsilon}+\frac{365}{3}\ln\frac{\Lambda}{4\pi T}+13\gamma_{E}
+\frac{236}{3}-\frac{110}{3}\frac{\zeta^{\prime}(-3)}{\zeta (-3)}\\
&&
+\frac{436}{3}\frac{\zeta^{\prime}(-1)}{\zeta (-1)}
\Big].
\eqa
Again, the sign $\approx$ indicates that the result is obtained using
strict perturbation theory.

Let us next consider ESQED. The contributions to 
$\ln {\cal Z}_{\mbox{\scriptsize ESQED}}$ 
comes from ordinary one.two and three-loop diagrams, as well as one-loop
graphs with one or two mass insertions.
The loop-integrals all vanish  
in the strict perturbation expansion, since there
is no mass scale. However, note that some of the three-loop diagrams
as well
the one-loop graphs with
two mass insertions and the two-loop graphs with a single mass insertion
are linearly divergent in the infrared. These divergences
are identical to those encountered in full SQED (the fifth and sixth
diagrams in Fig.~\ref{sqed3}). 

The vanishing of $\ln {\cal Z}_{\mbox{\scriptsize ESQED}}$ 
implies that the free energy in the strict perturbation
expansion is given by
\beq
-\frac{T\ln {\cal Z}}{V}\approx 
T\Big[f_{\mbox{\scriptsize ESQED}}(\Lambda)+
\delta f_{\mbox{\scriptsize ESQED}}(\Lambda)\Big].
\eeq
Here, $\delta f_{\mbox{\scriptsize ESQED}}(\Lambda)$ is the
counterterm of the unit operator.
According to~\cite{braaten}, $\delta f_{\mbox{\scriptsize ESQED}}(\Lambda)$ 
can be determined by calculating the 
logarithmic ultraviolet 
divergences in the effective
theory. Generally, $\delta f_{\mbox{\scriptsize ESQED}}(\Lambda)$ 
is given by a polynomial in the parameters of ESQED. 
At leading order it turns out that it is given by
\beq
\label{df}
\delta f_{\mbox{\scriptsize ESQED}}(\Lambda)=-\frac{e_{E}^2M^2}{2(4\pi )^2}
\frac{1}{\epsilon},
\eeq
which follows from a two-loop calculation in the next section.
Since the mass $M$ is multiplied by $1/\epsilon$, it is necessary
to expand it to first order in $\epsilon$ when expressing
$\delta f_{\mbox{\scriptsize ESQED}}(\Lambda)$ in terms of $e^2$, $\lambda$ and $T$. 
From~(\ref{skmass}) one finds
\beq
\frac{\partial M^2}{\partial\epsilon}\Bigg|_{\epsilon = 0}=
\frac{e^2T^2}{12}\Big[6\ln\frac{\Lambda}{4\pi T}+4+6\frac{\zeta^{\prime}(-1)}{\zeta (-1)}\Big]+
\frac{\lambda T^2}{18}\Big[2\ln\frac{\Lambda}{4\pi T}+2+2
\frac{\zeta^{\prime}(-1)}{\zeta (-1)}\Big].
\eeq
This implies that
\bqa\nonumber
\delta f_{\mbox{\scriptsize ESQED}}(\Lambda)T
&=&-\frac{e^4}{(4\pi )^2}\Big(\frac{T^2}{12}\Big)
\Big[\frac{18}{\epsilon}+36
\ln\frac{\Lambda}{4\pi T}+24+36\frac{\zeta^{\prime}(-1)}{\zeta (-1)}\Big]\\
&&-\frac{\lambda e^2}{(4\pi )^2}\Big(\frac{T^2}{12}\Big)
\Big[\frac{4}{\epsilon}+8
\ln\frac{\Lambda}{4\pi T}+8+8\frac{\zeta^{\prime}(-1)}{\zeta (-1)}\Big].
\eqa
Putting our results together, one finally obtains
\bqa\nonumber
\label{unit}
\!\!\!\!\!\!\!\!\hspace{-0.5cm}f_{\mbox{\scriptsize ESQED}}(\Lambda)T=
&=&
-\frac{2\pi^2T^4}{45}+\Big(\frac{T^2}{12}\Big)^2\Big[\frac{\lambda}{3}
+\frac{5e^2}{2}\Big]\\ \nonumber
&&-\frac{\lambda^2}{16\pi^2}\Big(\frac{T^2}{12}\Big)^2
\Big[\frac{10}{9}\ln\frac{\Lambda}{4\pi T}+\frac{4}{9}\gamma_{E}
+\frac{31}{45}
-\frac{2}{3}\frac{\zeta^{\prime}(-3)}{\zeta (-3)}
+\frac{4}{3}\frac{\zeta^{\prime}(-1)}{\zeta (-1)}
\Big]\\ \nonumber
&&-\frac{\lambda e^2}{16\pi^2}\Big(\frac{T^2}{12}\Big)^2
\Big[-4\ln\frac{\mu}{4\pi T}+16\ln\frac{\Lambda}{4\pi T}+4\gamma_{E}+\frac{20}{3}
+8\frac{\zeta^{\prime}(-1)}{\zeta (-1)}
\Big]\\ \nonumber
&&-\frac{e^4}{16\pi^2}\Big(\frac{T^2}{12}\Big)^2
\Big[\frac{41}{3}\ln\frac{\mu}{4\pi T}+72\ln\frac{\Lambda}{4\pi T}+13\gamma_{E}
+\frac{164}{3}-\frac{110}{3}\frac{\zeta^{\prime}(-3)}{\zeta (-3)}\\
&&
+\frac{328}{3}\frac{\zeta^{\prime}(-1)}{\zeta (-1)}
\Big].
\eqa
We have here used the renormalization group equations,
\bqa
\label{running}
\frac{de^2}{d\mu}&=&\frac{e^4}{24\pi^2},\\
\label{running2}
\frac{d\lambda}{d\mu}&=&\frac{5\lambda^2-18\lambda e^2+54e^4}{24\pi^2},
\eqa
to change the scale
from $\Lambda$ to $\mu$. The remaining logarithm of $\Lambda$
shows that $f_{\mbox{\scriptsize ESQED}}(\Lambda)$ depends explicitly on $\Lambda$. 
This logarithmic dependence on $\Lambda$ is necessary to cancel the 
$\Lambda$-dependence which arises in calculations in ESQED, as we shall see
in section IV.
This is contrast with spinor QED~\cite{jens}, where the unit operator is
independet of the factorization scale to order $e^4$.
\subsection{Renormalization Group Equations in ESQED}
Above, we have seen that some of the parameters in ESQED do not depend
explicitly on the cutoff $\Lambda$ to the order which we calculated 
them, while others do.
More generally, the coefficients in 
${\cal L}_{\mbox{\scriptsize ESQED}}$ satisfy a set of renormalization
group equations. The solutions to these equations can be used to sum up 
leading logarithms of the couplings constants from higher orders in the
perturbation expansion, as discussed in detail in Ref.~\cite{braaten}.
Although we shall not need these evolution equations in the present work,
we include a short discussion for completeness. They will be relevant
when calculations are pushed to higher orders.

The renormalization group equations follow from the requirement that physical
quantities such as the electric screening mass and the free energy be 
independent of this arbitrary factorization scale. If $C_n(\Lambda)$
denotes an arbitrary operator, one can write
\beq
\Lambda\frac{dC_n(\Lambda)}{d\Lambda}=\beta_n(C(\Lambda)).
\eeq
The beta functions $\beta_n(C(\Lambda))$ 
can written as power series in the coupling constants
of the effective theory.
Using dimensional arguments we can infer the general structure. Consider
first the coefficient of the unit operator $f_{\mbox{\scriptsize ESQED}}(\Lambda)$ and let us
restrict ourselves to the superrenormalizable terms which are explicitly
displayed in    ~(\ref{defleff}).
The unit operator has dimension three.
The beta function then contains terms in the form
$e^2_Em^2_E$,
$h_E^2m_E^2$, $\lambda_3m_E^2$,
$e^2_EM^2$,
$h_E^2M^2$ and $\lambda_3M^2$, 
as well a cubic polynomial in the three coupling constants $e^2_E$,
$h_E^2$ and $\lambda_3$. 
To order $\lambda e^2$, $\lambda e^2$ and $e^4$ only the first kind of terms
can appear. The coefficients in front of these terms are determined by 
calculating the logarithmic ultraviolet divergences in ESQED.
This is done in the next section, and the result is
\beq
\label{run1}
\Lambda\frac{df_{\mbox{\scriptsize ESQED}}(\Lambda)}{d\Lambda}=\frac{2e^2_EM^2}{(4\pi )^2}+{\cal O}(e^6).
\eeq
Using the fact that $e^2_E=e^2T$ at leading order and the leading order 
result for $M^2$, we can write the equation for 
$f_{\mbox{\scriptsize ESQED}}(\Lambda)$
in terms of $\lambda$ and $e^2$:
\beq
\Lambda\frac{df_{\mbox{\scriptsize ESQED}}(\Lambda)}{d\Lambda}=\frac{T^3}{(4\pi )^2}[\frac{e^4}{2}+\frac{\lambda e^2}{9}]+
{\cal O}(e^6).
\eeq
This form of the renormalization group equation could of course
have been obtained simply by differentiating~(\ref{unit}) with respect to 
$\Lambda$.

Consider next the mass parameter $m_E^2(\Lambda)$ and $M^2(\Lambda)$.
The beta functions for these parameters must be quadratic polynomials
in the $\lambda_3$, $e^2_E$ , $h^2_E$ and $\lambda_E$ (the latter is the 
coefficient in front of $\rho^2$).  
We have already
noted their independence of $\Lambda$ at leading order in the
couplings. 
In Ref.~\cite{jens2}, I have calculated $m^2_E(\Lambda)$ to
next-to-leading in $\lambda$ and $e^2$. The mass parameter turns out to be
independent of the factorization scale also to this order in the 
couplings:
\beq
\Lambda\frac{dm^2_E(\Lambda)}{d\Lambda}={\cal O}(e^6).
\eeq
This is in contrast with the corresponding result for the scalar mass parameter
$M^2(\Lambda)$. In Ref.~\cite{peisa}, 
this parameter has also been calculated at 
next-to-leading order in $\lambda$ and $e^2$, 
and the beta function reads
\beq
\label{run2}
\Lambda\frac{dM^2(\Lambda)}{d\Lambda}=\frac{1}{(4\pi )^2}
[6e^4_E-\frac{4\lambda_3 e^2_E}{3}+\frac{2\lambda^2_3}{9}]
+{\cal O}(e^6).
\eeq
Finally, we make the remark that the $\beta$ functions for the couplings
$\lambda_3(\Lambda)$,
$e^2_E(\Lambda)$, 
$h_E^2(\Lambda)$ and $\lambda_A(\Lambda)$ vanish
to order $\lambda^2$, $\lambda e^2$ and $e^4$~\cite{peisa}.
This property actually
holds to all orders in superrenormalizable interactions.

In summary, there are two parameters in ESQED whose evolution are
relevant up to order $e^6$. These are 
$f_{\mbox{\scriptsize ESQED}}(\Lambda)$ and $M^2(\Lambda)$. The
solutions to the renormalization group equations~(\ref{run1})
and~(\ref{run2}) are, respectively
\bqa
f_{\mbox{\scriptsize ESQED}}(\Lambda)
&=&f_{\mbox{\scriptsize ESQED}}(\Lambda^{\prime})
-\frac{2e^2_EM^2}{(4\pi)^2}
\ln\frac{\Lambda}{\Lambda^{\prime}},\\ 
M^2(\Lambda)
&=&M^2(\Lambda^{\prime})
+\frac{1}{(4\pi )^2}
[6e^4_E-\frac{4\lambda_3 e^2_E}{3}+\frac{2\lambda^2_3}{9}]
\ln\frac{\Lambda}{\Lambda^{\prime}}.
\eqa
\section{Calculations in ESQED}
\begin{figure}[htb]
\begin{center}
\mbox{\psfig{figure=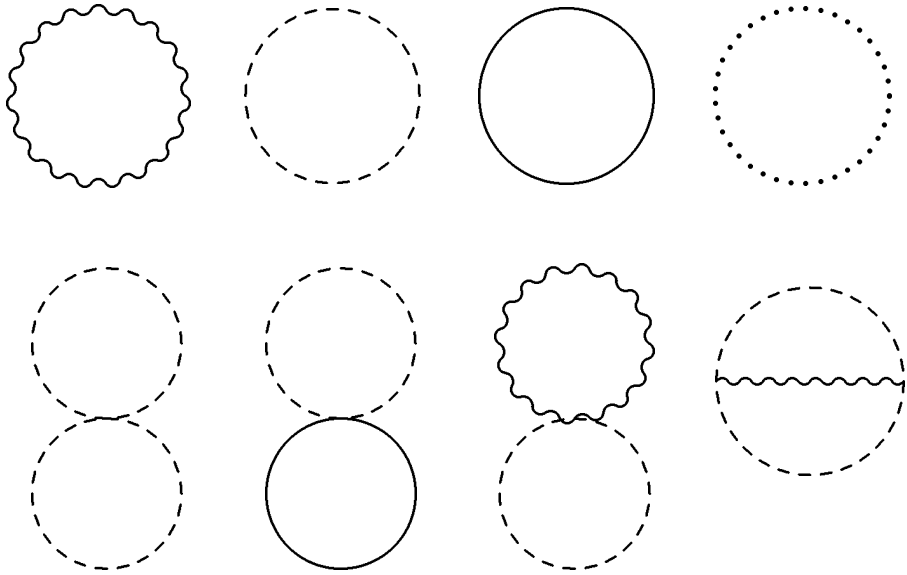}}
\end{center}
\caption[One and two-loop diagrams contributing to
the free energy in ESQED.]{\protect One and two-loop diagrams contributing to
the free energy in ESQED.}
\label{2lesqed}
\end{figure}

Having calculated the coefficients in ESQED to the order
required, we are now ready to calculate the free energy in SQED
to order $\lambda^2$, $\lambda e^2$ and $e^4$. The free energy is given by
the two terms in~(\ref{twoterm}), where 
$f_{\mbox{\scriptsize ESQED}}(\Lambda)T$ is given by~(\ref{unit}).
The remaining term $\ln{\cal Z}_{{\mbox{\scriptsize ESQED}}}$
is given by perturbative calculations using 
${\cal L}_{\mbox{\scriptsize ESQED}}$.
In order to obtain this term, it is
necessary that we take screening effects properly into account. 
This is done by including the mass parameters
$m_E^2(\Lambda)$ and $M^2(\Lambda)$ into the free part of the Lagrangian.
We then include the effects of the mass parameters to all orders, while
all other operators are treated as perturbations.
Thus, we split ${\cal L}_{\mbox{\scriptsize ESQED}}$. according to:
\bqa\nonumber 
({\cal L}_{\mbox{\scriptsize ESQED}})_{0}&=&\frac{1}{4}
F_{ij}F_{ij}+(\partial_{i}
\phi^{\dagger})(\partial_{i}\phi )+M^{2}(\Lambda )\phi^{\dagger}
\phi+\frac{1}{2}(\partial_{i}\rho)^{2}\\ \nonumber
&&
+\frac{1}{2}
m^{2}(\Lambda )\rho^{2}
+{\cal L}_{\mbox{\footnotesize gf}}
+{\cal L}_{\mbox{\footnotesize gh}}, \\ \nonumber
({\cal L}_{\mbox{\scriptsize ESQED}})_{\mbox{\footnotesize int}}&=&
e^{2}_{E}(\Lambda)\phi^{\dagger}\phi 
A_{i}^{3d}A_i^{3d}+h^2_E(\Lambda)\phi^{\dagger}\phi\rho^{2}
-ie_{E}(\Lambda )A_{i}^{3d}
(\phi^{\dagger}\partial_{i}\phi-\phi\partial_{i}\phi^{\dagger})\\
&&+\frac{\lambda_3(\Lambda)}{6} 
(\phi^{\dagger}\phi)^2
+\delta{\cal L}.
\eqa
The one and two-loop contributions to the pressure
are depicted in Fig.~\ref{2lesqed}. By adding the counterterm 
$\delta f_{\mbox{\scriptsize ESQED}}(\Lambda)$,
we obtain the contribution to the free energy from the momentum
scale $eT$.
\bqa\nonumber
\label{frieff}
-\frac{T\ln{\cal Z}_{{\mbox{\scriptsize ESQED}}}}{V}&=&
\frac{1}{2}T\int_p\ln (p^2+m^2_E)+T\int_p\ln (p^2+M^2)
+\frac{1}{2}(d-2)T\int_p\ln p^2\\ \nonumber
&&-\frac{1}{2}
e^2_{\footnotesize{E}}T\int_{pq}\frac{({\bf p}+{\bf q})^2}
{(p^2+M^2)(q^2+M^2)({\bf p}-{\bf q})^2}
+de^2_{\footnotesize{E}}T\int_{pq}\frac{1}{(p^2+M^2)q^2}\\ \nonumber
&&+e^2_{\footnotesize{E}}T\int_{pq}\frac{1}{(p^2+M^2)(q^2+m^2_E)}
+\frac{\lambda_3}{3}T\int_{pq}\frac{1}{(p^2+M^2)(q^2+M^2)}\\
&&+T\delta f_{\mbox{\scriptsize ESQED}}(\Lambda).
\eqa
Here, $d=3-2\epsilon$.
The two-loop contributions may be reduced to products of one-loop
integrals and a two-loop integral, which are tabulated in Appendix B.
This yields
\bqa\nonumber
-\frac{T\ln{\cal Z}_{{\mbox{\scriptsize ESQED}}}}{V}&=&
-\frac{m_E^3T}{12\pi}-\frac{M^3T}{6\pi}+
\frac{e_E^2M^2}{32\pi^2}+\frac{e^2_EM^2}{32\pi^2}
[\frac{1}{\epsilon}+2+4\ln\frac{\Lambda}{2M}]
+e^2_E\frac{Mm_E}{16\pi^2}\\
&&
+\frac{\lambda_3M^2}{48\pi^2}
+\delta f_{\mbox{\scriptsize ESQED}}(\Lambda).
\eqa
The logarithmic
ultraviolet divergence stems from the theta-diagram.
This pole in $\epsilon$ is then canceled by the
counterterm of the unit operator, 
$\delta f_{\mbox{\scriptsize ESQED}}(\Lambda)$, which is given by~(\ref{df}).
Using the expressions for the the parameters in ESQED
in terms of the couplings in full SQED as well as the factorization
scale $\Lambda$ and temperature,
we get the contribution to the free energy from the scale $eT$:
\bqa\nonumber
\label{energy2}
-\frac{T\ln{\cal Z}_{{\mbox{\scriptsize ESQED}}}}{V}&=&
-\frac{e^3T^4}{36\pi\sqrt{3}}-\frac{M^3T}{6\pi}+\frac{e^3MT^3}{16\pi^2\sqrt{3}}
+\frac{\lambda^2}{16\pi^2}\Big(\frac{T^2}{12}\Big)^2
\Big[\frac{8}{3}\Big]\\ 
&&+\frac{\lambda e^2}{16\pi^2}\Big(\frac{T^2}{12}\Big)^2
\Big[16\ln\frac{\Lambda}{M}+24
\Big]+\frac{e^4}{16\pi^2}\Big(\frac{T^2}{12}\Big)^2
\Big[72\ln\frac{\Lambda}{M}+54\Big].
\eqa
Adding~(\ref{unit}) and~(\ref{energy2}), we finally obtain the 
free energy of massless scalar electrodynamics at high temperature
through order $\lambda^2$, $\lambda e^2$ and $e^4$:
\bqa\nonumber
\label{main}
{\cal F}&=&-\frac{2\pi^2T^4}{45}
+\Big(\frac{T^2}{12}\Big)^2\Big[\frac{\lambda}{3}+\frac{5e^2}{2}\Big]
-\frac{e^3T^4}{36\pi\sqrt{3}}-\frac{M^3T}{6\pi}
+\frac{e^3MT^3}{16\pi^2\sqrt{3}}
\\ \nonumber
&&-\frac{\lambda^2}{16\pi^2}\Big(\frac{T^2}{12}\Big)^2
\Big[\frac{10}{9}\ln\frac{\Lambda}{4\pi T}+\frac{4}{9}\gamma_{E}
-\frac{89}{45}
-\frac{2}{3}\frac{\zeta^{\prime}(-3)}{\zeta (-3)}]
+\frac{4}{3}\frac{\zeta^{\prime}(-1)}{\zeta (-1)}
\Big]\\ \nonumber
&&-\frac{\lambda e^2}{16\pi^2}\Big(\frac{T^2}{12}\Big)^2
\Big[-4\ln\frac{\mu}{4\pi T}+16\ln\frac{2M}{4\pi T}
+4\gamma_{E}-\frac{52}{3}
+8\frac{\zeta^{\prime}(-1)}{\zeta (-1)}
\Big]\\ \nonumber
&&-\frac{e^4}{16\pi^2}\Big(\frac{T^2}{12}\Big)^2
\Big[\frac{41}{3}\ln\frac{\mu}{4\pi T}+72\ln\frac{2M}{4\pi T}
+13\gamma_{E}+\frac{2}{3}
-\frac{110}{3}\frac{\zeta^{\prime}(-3)}{\zeta (-3)}\\
&&
+\frac{328}{3}\frac{\zeta^{\prime}(-1)}{\zeta (-1)}
\Big].
\eqa
This represents the main result of the present paper, and
at this point some comments are in order.
Firstly, we note that the two-loop contribution in the gauge sector is
exactly the same as the one appearing in spinor QED~\cite{kast}. 
The terms which go like $e^3$ and $\lambda^{3/2}$ are non-analytic in the
coupling constants, and therefore demonstrate that one receives
contributions from all order in naive perturbation theory. The above
terms represent the leading order contributions from the infinite
string of infrared divergent ring diagrams.
We also note that the $\Lambda$-dependence cancels to next-to-leading
order in $\lambda$ and $e^2$, leaving a result which
is renormalization group
invariant to order $\lambda^2$, $\lambda e^2$ and $e^4$.
This can be easily checked
by using the renormalization
group equations,~(\ref{running}) and~(\ref{running2}), 
for the running coupling constants.

The effective field theory approach also explains the occurrence of logarithms
of $T/M$ in the expression for the free energy. We have seen that 
$f_{\mbox{\scriptsize ESQED}}(\Lambda)$ 
depends explicitly upon the factorization scale
$\Lambda$ and such terms appear as logarithms of $\Lambda/T$.
In ESQED, logarithms in the form $\Lambda/M$ occur in perturbative 
calculations.
In order to cancel the dependence of $\Lambda$ in the final answer,
these logarithms must match the logarithms from
$f_{\mbox{\scriptsize ESQED}}(\Lambda)$. 
We are then left with logarithms
of $M/T$. Hence these logarithms are connected to the renormalization of
$f_{\mbox{\scriptsize ESQED}}(\Lambda).$

A similar term $g^4\ln (m_E/T)$ has also been
 also found by Braaten and 
Nieto in the case of QCD~\cite{braaten3}.
No term of order $g^4\ln (m/T)$ arises
in $g^2\Phi^4$-theory, since $f(\Lambda)$ does not
run at next-to-leading order in $g^2$~\cite{braaten}.
This remark also applies to QED~\cite{jens}.
\section{Concluding Remarks}
In the present work we have calculated the free energy in SQED
to order
$\lambda^2$, $\lambda e^2$ and $e^4$
by effective field theory methods.
The major advantage of the effective field
theory approach is that it allows one to work with a single scale at a
time. This makes perturbative calculations simpler, since the scales $T$
and $eT$ are not intertwined in already complicated sum-integrals in
the full theory. However, we would also like to emphasize that 
effective field theory have some limitations. At present it is not clear
how to apply it in the computations of dynamical quantities. 
Here, the resummation program of Braaten and Pisarski or extensions
thereof are still essential ingredients of high temperature
field theory. 

We would also like to outline the calculations necessary to push the
calculations of the free energy to one more order in the coupling constants.
The coefficient of the unit operator are written as a power series in 
$\lambda$ and $e^2$, and so we already know this parameter to sufficient
precision. The free energy at one-loop in the effective theory goes like
$M^3(\Lambda)T$ and $m_E^3(\Lambda)T$. This implies that we need to know the
mass parameters in ESQED at next-to-leading order.
The former has been calculated by Farakos {\it et al.}
in Ref.~\cite{peisa} 
while the present author has
computed the latter in Ref.~\cite{jens2}. 
Furthermore, there are no new operators 
which contribute at this order, so all that
remains is calculating the three-loop diagrams in ESQED. 
As we have already noted both the coefficient of the unit operator and
the scalar mass parameter depend explicitly on the scale $\Lambda$.
We realize that, upon expanding $M^2(\Lambda)$ in powers of coupling
constants, we get terms proportional to $e^5\ln\Lambda /T$.
This allows us to predict terms in the form $e^5\ln M/T$ or 
$e^5\ln m_E/T$ 
in the
pressure of scalar electrodynamics. 
This also explains the absence of such terms
in QCD~\cite{braaten3} as well
as QED~\cite{jens}, since the corresponding
mass parameter $m_E^2(\Lambda)$ is independent of $\Lambda$ 
to fourth order in the gauge coupling.

Finally, we would like to briefly
comment upon other applications of the present 
approach. 
This method is perhaps the most transparent way of doing effective
field theory at finite temperature, and it would certainly
be worthwhile extending it to e.g. the study of 
phase transtions. 
This aproach would then
represent an alternative to the methods
which explicitly construct the effective field theory of the light mode
by integrating over the heavy modes~\cite{laine}.
Work along these lines is in progress~\cite{prep}.

\section*{Acknowledments}
The author would like to thank Eric Braaten and Finn Ravndal for  
valuable discussions.
\appendix\bigskip\renewcommand{\theequation}{\thesection.\arabic{equation}}
\setcounter{equation}{0}\section{Sum-integrals in the Full Theory}
In this appendix we give the necessary details for
the sum-integrals used in present work. 
Throughout the work we use the imaginary 
time formalism, where the four-momentum is $P=(p_{0},{\bf p})$
with $P^{2}=p_{0}^{2}+{\bf p}^{2}$. 
The Euclidean energy takes on discrete values, $p_{0}=2n\pi T$
for bosons.
Dimensional regularization is used to
regularize both infrared and ultraviolet divergences by working
in $d=4-2\epsilon$ dimensions, 
and we apply the $\overline{\mbox{MS}}$ 
renormalization scheme. 
We use the following shorthand notation
for the sum-integrals that appear below:
\bqa
\hbox{$\sum$}\!\!\!\!\!\!\int_P f(P)&\equiv 
&\Big( \frac{e^{\gamma_{\tiny E}}\mu^{2}}
{4\pi}\Big )^{\epsilon}\,\,\,\,
T\!\!\!\!\!
\sum_{p_{0}=2\pi nT}\int\frac{d^{3-2\epsilon}k}
{(2\pi)^{3-2\epsilon}}f(P).
\eqa
Then $\mu$ coincides with the
renormalization scale in the $\overline{\mbox{MS}}$ renormalization
scheme.

Arnold and Zhai~\cite{arnold1} 
have developed a new machinery to deal with complicated
sum-integrals. The methods are completely analytic and represent 
significant progress in perturbative calculations at finite temperature.
They have calculated and listed all the sum-integrals needed in the present
work, except for the one in~(\ref{bcal}). This sum-integral
has been evaluated by Braaten and Nieto~\cite{braaten} using the methods
of Ref.~\cite{arnold1}. We reproduce them below for the convenience of
the reader.\\ \\The specific one-loop sum-integrals needed are
\bqa
\hbox{$\sum$}\!\!\!\!\!\!\int_P\ln P^{2}&=&-\frac{\pi^{2}T^{4}}{45}\Big[1
+O(\epsilon )\Big], \\ 
\label{1l1}
\hbox{$\sum$}\!\!\!\!\!\!\int_P\frac{1}{P^{2}}&=&\frac{T^{2}}{12}
\Big[1+\Big(2\ln\frac{\mu}{4\pi T}+
2+2\frac{\zeta^{\prime}(-1)}{\zeta (-1)}\Big)\epsilon
+O(\epsilon^{2})\Big],\\
\label{1l2}
\hbox{$\sum$}\!\!\!\!\!\!\int_P\frac{1}{(P^{2})^{2}}&=&\frac{1}{(4\pi )^{2}}
\Big[\frac{1}{\epsilon}+2\ln\frac{\mu}{4\pi T}
+2\gamma_{E}+O(\epsilon )\Big],\\
\label{1l3}
\hbox{$\sum$}\!\!\!\!\!\!\int_P\frac{p_{0}^{2}}{(P^{2})^{2}}
&=&-\frac{T^{2}}{24}
\Big[1+\Big(2\ln\frac{\mu}{4\pi T}
+2\frac{\zeta^{\prime}(-1)}{\zeta (-1)}\Big)\epsilon
+O(\epsilon^{2} )\Big].
\eqa
All of the two-loop sum-integrals encountered in this work factorize into
products of the one-loop sum-integrals listed above. 
The only
two-loop sum-integral needed is:
\bqa
\label{zweiloop}
\hbox{$\sum$}\!\!\!\!\!\!\int_{PQ}\frac{1}{P^{2}Q^{2}(P+Q)^{2}}&=&0.
\eqa
Some of the three-loop sum-integrals factorize into products of three one-loop
sum-integrals, while others factorize into a product of two one-loop
sum-integrals and the two-loop sum-integral in~(\ref{zweiloop}).
The specific three-loop sum-integrals needed in the present work are
\bqa\nonumber
\hbox{$\sum$}\!\!\!\!\!\!\int_{PKQ}\frac{1}{P^2Q^2K^2(P+K+Q)^2}&=&
\frac{1}{(4\pi )^2}\frac{T^4}{144}\Big[\frac{6}{\epsilon}+36\ln\frac{\mu}{4\pi T}-12\frac{\zeta^{\prime}(-3)}{\zeta (-3)} \\
&&+48\frac{\zeta^{\prime}(-1)}{\zeta (-1)}+\frac{182}{5}\Big]+O(\epsilon )\\\nonumber
\label{bcal}
\hbox{$\sum$}\!\!\!\!\!\!\int_{PKQ}\frac{(P-Q)^4}{P^2Q^2K^2(P+K+Q)^2}&=&
\frac{1}{(4\pi )^2}\frac{T^4}{144}\Big[\frac{22}{3\epsilon}+\frac{73}{3}
+8\gamma_E
+\frac{64}{3}\frac{\zeta^{\prime}(-1)}{\zeta (-1)}-\\
&&
\frac{20}{3}\frac{\zeta^{\prime}(-3)}{\zeta (-3)} 
\Big]+O(\epsilon ).
\eqa
The first of these three-loop sum-integrals is the bosonic 
basketball~\cite{arnold1}, while the second is denoted by ${\cal M}_{2,-2}$
in Ref~\cite{braaten3}. 
\setcounter{equation}{0}\section{Integrals in the Effective Theory}
In the effective three-dimensional theory we use dimensional regularization
in $3-2\epsilon$ dimensions to regularize infrared and ultraviolet 
divergences.
In analogy with Appendix A, we define
\beq
\int_{p}f(p)\equiv\Big( \frac{e^{\gamma_{\tiny E}}\mu^{2}}
{4\pi}\Big )^{\epsilon}\int\frac{d^{3-2\epsilon}p}
{(2\pi)^{3-2\epsilon}}f(p).
\eeq
Again $\mu$ coincides with the renormalization scale in the 
modified minimal subtraction renormalization
scheme.\\ \\
All the integrals needed in the present work have been evaluated and
listed by Braaten and Nieto in Ref.~\cite{braaten}. We list them below
for the convenience of the reader.\\ \\
In the effective theory we need the following one-loop integrals
\bqa
\int_{p}\ln (p^{2}+m^{2})&=&-\frac{m^{3}}{6\pi}
\Big[1+\epsilon\Big(2\ln\frac{\mu}{2m}+\frac{8}{3}\Big)
+O(\epsilon^{2})\Big ], \\
\int_{p} \frac{1}{p^{2}+m^{2}}&=&-\frac{m}{4\pi}
\Big[1+\Big(2\ln\frac{\mu}{2m}+2\Big)\epsilon
+O(\epsilon^{2})\Big],\\
\int_{p} \frac{1}{(p^{2}+m^{2})^{2}}&=&\frac{1}{8\pi m}
\Big[1+\Big(2\ln\frac{\mu}{2m}\Big)\epsilon
+O(\epsilon^{2})\Big].
\eqa
The two-loop integrals in the effective theory
also factorize into products of one-loop integrals plus 
a single two-loop integral. More, specifically one needs
\bqa\nonumber
\int_{pq} \frac{1}{(p^{2}+m^{2})(q^{2}+m^{2})({\bf p}-{\bf q})^{2}}&=&
\frac{1}{(4\pi )^{2}}
\Big[\frac{1}{4\epsilon}
+\frac{1}{2}
+\ln\frac{\mu}{2m}+O(\epsilon )\Big].
\eqa

\pagebreak


\end{document}